%% file: main.tex
\documentclass[sigconf]{acmart}
\AtBeginDocument{%
  \providecommand\BibTeX{{%
    \normalfont B\kern-0.5em{\scshape i\kern-0.25em b}\kern-0.8em\TeX}}}

\copyrightyear{2022} 
\acmYear{2022} 
\setcopyright{acmlicensed}\acmConference[ASE '22]{37th IEEE/ACM International Conference on Automated Software Engineering}{October 10--14, 2022}{Rochester, MI, USA}
\acmBooktitle{37th IEEE/ACM International Conference on Automated Software Engineering (ASE '22), October 10--14, 2022, Rochester, MI, USA}
\acmPrice{15.00}
\acmDOI{10.1145/3551349.3556897}
\acmISBN{978-1-4503-9475-8/22/10}

\usepackage[ruled,linesnumbered]{algorithm2e}
\usepackage{algorithmicx}
\usepackage{hyperref}

\usepackage{multirow}
\usepackage{booktabs}
\usepackage{tabularx}
\usepackage{enumitem}
\usepackage{caption}
\usepackage{verbatim}
\usepackage{todonotes}
\usepackage[normalem]{ulem}

\usepackage{listings}
\usepackage{xcolor}

\newcommand{\mt}[1]{\text{\tt{#1}}}

\definecolor{darktangerine}{rgb}{1.0, 0.66, 0.07}
\definecolor{darkspringgreen}{rgb}{0.09, 0.45, 0.27}
\definecolor{codegreen}{rgb}{0,0.6,0}
\definecolor{codegray}{rgb}{0.5,0.5,0.5}
\definecolor{codepurple}{rgb}{0.58,0,0.82}
\definecolor{backcolour}{rgb}{0.95,0.95,0.92}

\lstdefinestyle{mystyle}{
    backgroundcolor=\color{backcolour},   
    commentstyle=\color{codegreen},
    keywordstyle=\color{magenta},
    numberstyle=\tiny\color{codegray},
    stringstyle=\color{codepurple},
    basicstyle=\ttfamily\footnotesize,
    breakatwhitespace=false,         
    breaklines=true,                 
    captionpos=b,                    
    keepspaces=true,                 
    numbers=left,                    
    numbersep=5pt,                  
    showspaces=false,                
    showstringspaces=false,
    showtabs=false,                  
    tabsize=2
}

\newtheorem{theorem}{Theorem}[section]

\newtheorem{proposition}[theorem]{Proposition}
\newtheorem{example}[theorem]{Example}

\newcommand{\coolname}{$\mathtt{LawBreaker}$\xspace}

\begin{document}

\title{LawBreaker: An Approach for Specifying Traffic Laws and Fuzzing Autonomous Vehicles}

 \author{Yang Sun}
 \orcid{0000-0002-2409-2160}
 \affiliation{%
   \institution{Singapore Management University}
   \country{Singapore}
}
\email{yangsun.2020@phdcs.smu.edu.sg}

\author{Christopher M. Poskitt}
\orcid{0000-0002-9376-2471}
\affiliation{\institution{Singapore Management University}\country{Singapore}}
\email{cposkitt@smu.edu.sg}

\author{Jun Sun}
\orcid{0000-0002-3545-1392}
\affiliation{\institution{Singapore Management University}\country{Singapore}}
\email{junsun@smu.edu.sg}

 \author{Yuqi Chen}
 \affiliation{%
   \institution{ShanghaiTech University}
   \country{China}
   }
 \email{chenyq@shanghaitech.edu.cn}

 \author{Zijiang Yang}
 \affiliation{%
  \institution{GuardStrike Inc.}
  \country{China}
  }

\input{Abstract}



\maketitle

\input{Introduction}

\input{Overview_Background}

\input{Language_Description}

\input{Fuzzing}

\input{Evaluation}

\input{RelatedWork}

\input{Conclusion}

\bibliographystyle{ACM-Reference-Format}
\balance
\bibliography{reference}

\end{document}

%% file: Abstract.tex
\begin{abstract}
	Autonomous driving systems~(ADSs) must be tested thoroughly before they can be deployed in autonomous vehicles.
	High-fidelity simulators allow them to be tested against diverse scenarios, including those that are difficult to recreate in real-world testing grounds.
	While previous approaches have shown that test cases can be generated automatically, they tend to focus on weak oracles (e.g.~reaching the destination without collisions) without assessing whether the journey itself was undertaken safely and satisfied the law.
	In this work, we propose \coolname, an automated framework for testing ADSs against real-world traffic laws, which is designed to be compatible with different scenario description languages.
	\coolname provides a rich driver-oriented specification language for describing traffic laws, and a fuzzing engine that searches for different ways of violating them by maximising specification coverage.
	To evaluate our approach, we implemented it for Apollo+LGSVL and specified the traffic laws of China.
    \coolname was able to find 14 violations of these laws, including 173 test cases that caused accidents.
\end{abstract}

%% file: Introduction.tex
\section{Introduction}
Autonomous driving systems~(ADSs) combine sensors and software to control, navigate, and drive autonomous vehicles~(AVs).
As inherently safety-critical systems, ADSs must be comprehensively tested before they can be deployed on public roads.
High-fidelity simulators (e.g.~LGSVL~\cite{rong2020lgsvl}, CARLA~\cite{dosovitskiy2017carla}) play an important role in this effort as they allow ADSs to be evaluated across a broad range of scenarios.
This includes scenarios that are hard to recreate in real-world testing grounds, but are important to evaluate since an incorrect decision by the ADS could lead to an accident~\cite{favaro2017examining,dixit2016autonomous}.

In black box simulator-based testing, the ADS is systematically evaluated against a number of different \emph{scenarios} and \emph{oracles}.
Scenarios are configurations of objects on a map (e.g.~obstacles, pedestrians, and vehicles) as well as their dynamic behaviour, and can be described to different degrees by domain-specific languages~(DSLs) such as Scenic~\cite{fremont2019scenic}, CommonRoad~\cite{althoff2017commonroad}, GeoScenario~\cite{queiroz2019geoscenario}, and AVUnit~\cite{AVUnit2021}.
Oracles are `pass/fail' criteria that the ADS must satisfy under every test scenario~\cite{neurohr2020fundamental}.
Unfortunately, existing testing frameworks tend to use weak oracles. AV-Fuzzer~\cite{li2020av}, for example, evaluates ADSs on their ability to complete a journey without getting too close to other vehicles.
Criteria based on getting from A to B without collisions are no doubt important, but for AVs, the journey is as important as the destination, and we need richer criteria about how an AV undertakes it.
Jumping red lights at every junction is clearly unacceptable, for example, even if the ADS manages to achieve it without collisions.

Fortunately, rich sets of criteria for how a vehicle should undertake a journey already exist: the various national \emph{traffic laws}.
In addition to avoiding collisions, an ADS should satisfy the traffic laws of the country it operates in. Until we design new traffic laws specifically for ADSs, existing traffic laws remain the gold standard for ensuring road safety.
Testing an ADS against such traffic laws, however, is challenging.
First, they are typically expressed in natural language with respect to the driver's perspective.
This leads to non-intuitive encodings in existing specification languages that are based on a global view~(e.g.~\cite{AVUnit2021}).
Second, traffic laws vary across countries, so a general and adaptable specification language is necessary (instead of a fixed built-in oracle for one country).
Unfortunately, existing specification approaches for traffic laws have limited reusability and extensibility.
For example, rulebooks~\cite{censi2019liability, collin2020safety} focus on the logic transition process and do not provide a natural way to describe laws, whereas other formalisations (e.g.~\cite{rizaldi2015formalising, esterle2020formalizing, maierhofer2020formalization, rizaldi2017formalising}) are tightly coupled with the test scenarios, i.e.~the laws must be customised for each new scenario.

\begin{figure}[t]
    \centering
    \includegraphics[width=0.9\linewidth]{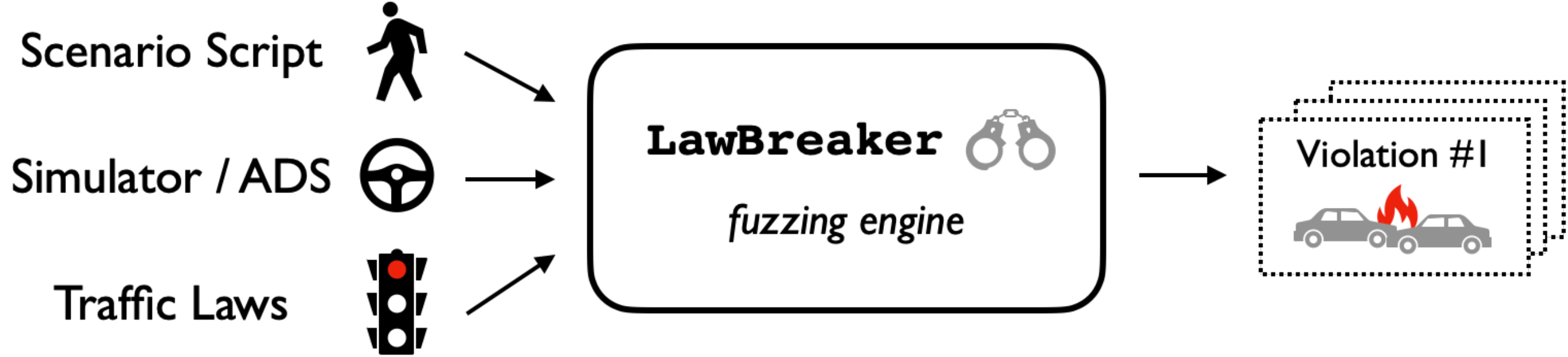}
    \caption{High-level workflow of \coolname}
    \label{fig:workflow}
\end{figure}

In this work, we present the design and implementation of \coolname, a DSL for specifying traffic laws and an automated framework for testing ADSs against them.
First, our language allows users to specify traffic laws more naturally from the perspective of the driver (instead of globally), e.g. ``at a junction the driver should give way to pedestrians when turning right''.
Second, \coolname is decoupled from any particular testing scenario, i.e.~not only can the same laws be interpreted across different maps, but they can be used together with any DSL for generating test scenarios (i.e.~placements of vehicles, pedestrians, and obstacles in a map).
Finally, \coolname provides a fuzzing engine that searches for different violations of laws by attempting to cover as many different ways of violating the specification as possible.
These uncovered violations may then provide clues on how to improve ADSs.

The workflow of \coolname is summarised in Figure~\ref{fig:workflow}. Users provide a scenario script, an ADS, a simulator, and some traffic laws specified using our language. Our fuzzing engine then systematically generates test cases in the simulator that try to cause the ADS to violate those laws, revealing flaws to be addressed in the design of the ADS. These violations are recorded, and can be played back visually by using the simulator.

Our implementation of \coolname consists of:
(1)~a grammar parser, which uses antlr4~\cite{antlr4}, to extract the elements describing a scenario and the corresponding traffic laws;
(2)~a fuzzing engine, that implements our specification-coverage guided fuzzing algorithm;
and (3)~a bridge, which connects the grammar parser, fuzzing engine, ADSs, and simulator to make the whole system run.
We evaluate this implementation using AVUnit~\cite{AVUnit2021} for scenario scripts, LGSVL~\cite{rong2020lgsvl} as the simulator, and different versions of Baidu Apollo~\cite{apollo60, apollo50} as the ADSs under test. As Apollo was designed by a Chinese company, we chose to evaluate it under Chinese traffic laws.
In particular, we specified and tested 24 Chinese traffic laws in \coolname, finding that 14 of them were violated by Apollo, and that 173 of the test cases generated also caused accidents. Videos of some of these violations can be found online \cite{ourweb}.

%% file: Overview_Background.tex
\section{Overview of LawBreaker}
\label{sec:overview_and_background}

\begin{figure}[!t]
    \centering
    \includegraphics[width=1\linewidth]{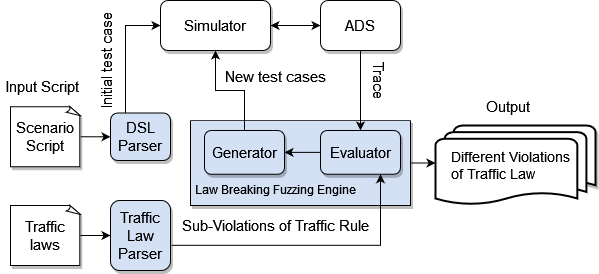}
    \caption{The architecture of \coolname}
    \label{fig:architecturel}
    \vspace{-0.5cm}
\end{figure}

The overall architecture of \coolname and how it interfaces with existing simulators and ADSs is shown in Figure~\ref{fig:architecturel}.
It has three main components: an existing DSL for describing scenes and scenarios (AVUnit~\cite{AVUnit2021}); our new driver-oriented specification language for traffic laws, based on signal temporal logic (Section~\ref{sec:Spec}); and the \coolname fuzzing algorithm (Section~\ref{sec:Fuzzing}).
Our architecture is fully decoupled, and intended to be compatible with different ADSs (e.g.~different versions of Apollo~\cite{apollo60} and Autoware~\cite{autoware}).

First, the scenario script component prepares the initial test case of the simulator by translating the specified scenario into the required API calls.
Second, the traffic law component describes testing oracles from the driver's view without needing any particular knowledge about the map or its agents.
Finally, the fuzzing engine repeatedly extracts a trace from the ADS, evaluates it against the specification, and uses the outcome to generate new test cases for the simulator to run.
Note that this algorithm uses the grammar of the scenario DSL when generating new testing scenarios.
The simulator itself (e.g.~LGSVL or CARLA) treats the ADS as a black box.
When tests are underway, the ADS extracts sensory information of the AV from the simulator.
This includes data from perception equipment (e.g.~camera, Lidar, and GPS) and chassis control (e.g.~brake, gas, and steer).

In order to specify testing scenarios, we utilise an existing DSL called AVUnit~\cite{AVUnit2021}.
AVUnit specifies the motions of NPC vehicles and pedestrians (i.e.~non-player characters representing objects other than the AV under test), and other environment-related information such as time and weather. AVUnit is a highly expressive language which allows us to specify detailed scenarios, e.g.~the status of every NPC vehicle as well as their trajectories. We refer the readers to~\cite{AVUnit2021} for details on AVUnit and remark that alternative scenario DSLs (e.g.~Scenic~\cite{fremont2019scenic}) can be adopted for \coolname easily. 
Note that even though AVUnit describes the motion task of the ego vehicle, the trajectory of the ego vehicle is determined by the ADS. \\

\begin{figure}[!t]
\setlength{\abovecaptionskip}{0.cm}
\setlength{\belowcaptionskip}{-5.cm}
\begin{lstlisting}[caption={Specifying Article \#38 in \coolname}, label={lst:spec}, language=C++,
    style=myStyle,escapechar=@]
Trace trace = EXE(scenario0);
// Green Lights
law38_sub1_1 = (trafficLightAhead.color == green) & (stoplineAhead(2) | junctionAhead(2)) & ~PriorityNPCAhead & ~PriorityPedsAhead;
law38_sub1_2 = F[0,2](speed > 0.5);
law38_sub1 = G (law38_sub1_1 -> law38_sub1_2); 

// Yellow Lights
law38_sub2_1 = ((trafficLightAhead.color == yellow) & (stoplineAhead(0) | currentLane.number == 0)) -> (F[0,2] (speed > 0.5));
law38_sub2_2 = ((trafficLightAhead.color == yellow) & stoplineAhead(3.5) & ~stoplineAhead(0) & currentLane.number > 0) -> (F[0,3] (speed < 0.5));
law38_sub2 = G (law38_sub2_1 & law38_sub2_2); @\label{line:law38_sub2}@

// Red Lights
law38_sub3_1 = ((trafficLightAhead.color == red) & (stoplineAhead(2) | junctionAhead(2)) & ~(direction == right)) -> (F[0,3] (speed < 0.5));
law38_sub3_2 = ((trafficLightAhead.color == red) & (stoplineAhead(2) | junctionAhead(2)) & direction == right & ~PriorityNPCAhead & ~PriorityPedsAhead) -> (F[0,2] (speed > 0.5));
law38_sub3 = G (law38_sub3_1 & law38_sub3_2); 

law38 = law38_sub1 & law38_sub2 & law38_sub3;
trace |= law38;
\end{lstlisting}
\vspace{-0.5cm}
\end{figure}

\noindent \textbf{\emph{Illustrative Example.}}
Listing~\ref{lst:spec} presents an example of a traffic law specification in \coolname.
In particular, it describes Article \#38 of the \emph{Regulations for the Implementation of the Road Traffic Safety Law of the People's Republic of China}~\cite{China_traffic_law}, which stipulates how a vehicle should behave with respect to a traffic light at a junction.
An English translation of Article \#38 reads as follows:
\emph{
\begin{enumerate}
    \item[(1)] When the \textcolor{darkspringgreen}{green} light is on, vehicles are allowed to pass, but turning vehicles shall not hinder the passing of vehicles going straight and pedestrians who are crossing;
    \item[(2)] When the \textcolor{darktangerine}{yellow} light is on, vehicles that have crossed the stop line can continue to pass;
    \item[(3)] When the \textcolor{red}{red} light is on, vehicles are prohibited from passing. However, vehicles turning right can pass without hindering the passage of vehicles or pedestrians.
\end{enumerate}
}

\noindent Article \#38 is specified as $\mathtt{law38}$ in Listing~\ref{lst:spec}, which in turn consists of three conjuncts separately describing clauses (1)--(3).
In this example, we focus on the rules concerning yellow lights in (2), which are specified as $\mathtt{law38\_sub2}$ (Line~\ref{line:law38_sub2}).
Note that $\mathtt{law38\_sub2}$ is a temporal expression (indicated by `$\mathtt{G}$' for `\emph{always}'), which indicates that its two parts $\mathtt{law38\_sub2\_1}$ and $\mathtt{law38\_sub2\_2}$ should always be satisfied by the vehicle under test.

First, $\mathtt{law38\_sub2\_1}$ specifies that if the traffic light is yellow and the vehicle is on the stop line, then the vehicle should proceed through the junction.
Article \#38 does not state how quickly the car should move through the junction in this scenario: like most traffic laws, there is a degree of ambiguity that is expected to be resolved by common sense or practice.
In \coolname, however, we need to be more precise so that our specification can serve as an oracle, so we interpret the law as requiring that the vehicle will `\emph{eventually}' ($\mathtt{F}$) move within the next $d$ time steps. The variable $d$ can be customised by the user; we set $d$ to 2 here.

Second, $\mathtt{law38\_sub2\_2}$ specifies that if the traffic light is yellow and the vehicle is a safe distance away from the stop line, then the vehicle is expected to eventually (`$\mathtt{F}$') stop within the next three time steps.
The laws regarding green ($\mathtt{law38\_sub1}$) and red ($\mathtt{law38\_sub3}$) lights are defined in a similar manner using the temporal and Boolean operators of \coolname.

Ultimately, the goal of \coolname is to be able to automatically generate test scenarios in which these specifications are \emph{violated}.
Furthermore, it aims to find as many different ways of violating the laws as possible, e.g.~by finding different counterexamples for each of the sub-expressions (e.g.~$\mathtt{law38\_sub2\_1}$). To illustrate this, consider the following two test cases generated by \coolname.
    In the first test case, the AV already reached the stop line when the traffic light turned from green to yellow, but hesitated for long enough that it actually crosses the intersection during a red light.  
    Hence, the AV violates not only the $\mathtt{law38\_sub2\_1}$ but also $\mathtt{law38\_sub3\_1}$.
    In the second test case, the AV rushed the yellow light and \emph{caused an accident}.
    In this situation, the AV violated $\mathtt{law38\_sub2\_2}$ by rushing through even though there was enough distance to stop.

%% file: Language_Description.tex
\section{Specifying Traffic Laws}
\label{sec:Spec}
We introduce the syntax and semantics of our DSL for specifying driver-oriented traffic laws and present our main case study.

\begin{figure}
\footnotesize
\begin{align*}
Formula~~~\longrightarrow & \quad BoolExpr \;|\; \mt{'}\!\sim\!\mt{'} Formula\;|\; Formula\ \mt{'}\mathtt{\&}{'}\ Formula\;| \\
&\quad Formula\ \mt{'|'}\ Formula\;|\; Formula\ \mt{'U'}\ Interval\ Formula\;| \\
&\quad \mt{'G'}\ Interval\ Formula\;|\; \mt{'F'}\ Interval\ Formula\ |\ \mt{'N'}\ Formula \\[1mm]
BoolExpr~~~\longrightarrow & \quad x_b \; | \; e_1 \oplus e_2
\end{align*} 
    \caption{Syntax of \coolname formulas, where $x_b$ is a Bool variable; $e_1$ and $e_2$ are expressions; and $\oplus$ is a relational operator, i.e.~$\oplus \in \{ =, >, <, \leq, \geq\}$}
    \label{fig:syntax of language}
\end{figure}

\subsection{Syntax and Semantics}
\label{sec:syntax_semantics}
We first need the concept of a \emph{trace} for understanding the syntax of our specification language.
A trace is a sequence of scenes, and a scene is a snapshot of the world (i.e.~the status of all vehicles, pedestrians, and so on). 
In \coolname, we offer a range of variables which allow us to extract relevant information from the scene, forming the building blocks of our specification language.

At the top-level, our specification language takes the form of temporal logic formulas.
The syntax is shown in Figure~\ref{fig:syntax of language}, where $\mathtt{U}$, $\mathtt{G}$, $\mathtt{F}$, and $\mathtt{N}$ respectively represent the temporal operators `until', `always', `eventually', and `next'.
Furthermore, $Interval$ is a real-time interval $\mt{[}a\mathtt{,}\ b\mt{]}$ in which $a$ and $b$ are numerical expressions.

Intuitively, the formula $f_1\ \mathtt{U}\mt{[}\mathtt{a, b}\mt{]}\ f_2$ (with $f_1, f_2$ derived from $Formula$) expresses that $f_1$ is true at every time point $t$ of the trace until $f_2$ becomes true within the time interval $[t+a, t+b]$.
Similarly, $\mathtt{G}\mt{[}\mathtt{a, b}\mt{]} f_1$ (resp.~$\mathtt{F}\mt{[}\mathtt{a, b}\mt{]} f_1$) is true if the formula $f_1$ always (resp.~eventually) holds in time interval $[t+a, t+b]$ for every time point $t$.
Finally, $\mathtt{N}\ f_1$ holds if $f_1$ is true in the next time point.

Temporal formulas are defined over Boolean expressions, which are built over several domain-specific variables related to the driver and its immediate surroundings.
For example, the Bool variable $\mathtt{isTrafficJam}$ is true if there is a traffic jam ahead of the AV under test.
Using this variable is more convenient, for example, than a `global' approach which would require some quantification over all other NPC vehicles and a judgement based on the direction of travel and the position of the AV.
As there are a range of variables in our language, we organise them into a few categories. \\

\begin{table}[!t]
	\centering
	\caption{Car status variables in \coolname}
	\label{tab:car_status_variables}
	\footnotesize
    \begin{tabular}{c|c|p{1.5in}}
         Variable & Type & Remarks  \\
         \hline
         $\mathtt{highBeamOn}$ & Bool & -- \\ 
         $\mathtt{lowBeamOn}$ & Bool & -- \\
         $\mathtt{turnSignal}$ & Enum & $\mathtt{off}$, $\mathtt{left}$, or $\mathtt{right}$ \\
         $\mathtt{fogLightOn}$ & Bool & APIs not available in Apollo \\
         $\mathtt{hornOn}$ & Bool & -- \\
         $\mathtt{warningFlashOn}$ & Bool & APIs not available in Apollo \\
         $\mathtt{gear}$ & Enum & $\mathtt{NEUTRAL}$, $\mathtt{DRIVE}$, $\mathtt{REVERSE}$, $\mathtt{PARK}$, $\mathtt{LOW}$, $\mathtt{INVALID}$, or $\mathtt{NONE}$ \\
         $\mathtt{engineOn}$ & Bool & -- \\
         $\mathtt{direction}$ & Enum & $\mathtt{forward}$, $\mathtt{left}$, $\mathtt{right}$ \\
         $\mathtt{toManual}$ & Bool & True if and only if AV control passed to human operator \\
    \end{tabular}
\end{table}

\noindent \textbf{\emph{Car and Driving Status Variables.}}
Car status variables can be used to describe properties involving the lights, engine, horn, and direction of the AV.
The properties supported by \coolname are summarised in Table~\ref{tab:car_status_variables}.
These variables are self-explanatory (e.g.~$\mathtt{hornOn}$ is true if and only if the horn is sounding) and are either of Bool or enumerated type.

\begin{table}[!t]
	\centering
	\caption{Driving status variables in \coolname}
	\label{tab:driving_status_variables}
	\footnotesize
    \begin{tabular}{c|c|p{1.7in}}
         Variable & Type & Remarks  \\
         \hline
         $\mathtt{speed}$ & Number & Speed of ego vehicle (km/h) \\ 
         $\mathtt{acc}$ & Number & Acceleration of ego veh.~(m/s$^2$) \\
         $\mathtt{brake}$ & Number & Braking percentage of ego veh.~(\%)\\
         $\mathtt{isChangingLane}$ & Bool & -- \\
         $\mathtt{isOverTaking}$ & Bool & -- \\
         $\mathtt{isTurningAround}$ & Bool & -- \\
    \end{tabular}
\end{table}

Driving status variables can be used to describe the speed, acceleration, and braking status of the AV.
Furthermore, there are Bool variables that capture manouevres that the AV is currently undertaking, e.g.~changing lanes, overtaking another vehicle, or turning around.
Table~\ref{tab:driving_status_variables} summarises them. \\

\begin{table}[!t]
	\centering
	\caption{Road variables in \coolname}
	\label{tab:road_variables}
	\footnotesize
    \begin{tabular}{c|c|p{0.9in}}
         Variable & Type & Remarks  \\
         \hline
         $\mathtt{currentLane}$ & Lane & -- \\
         $\mathtt{speedLimit}$ & SpeedLimit & -- \\ 
         $\mathtt{streetLightOn}$ & Bool & -- \\
         $\mathtt{honkingAllowed}$ & Bool & -- \\ 
         $\mathtt{crosswalkAhead(n)}$ & Bool & Within distance $\mathtt{n}$ \\
         $\mathtt{junctionAhead(n)}$ & Bool & Within distance $\mathtt{n}$ \\ 
         $\mathtt{specialLocationAhead(n)}$ & SpecialLocation & Within distance $\mathtt{n}$ \\ 
         $\mathtt{stoplineAhead(n)}$ & Bool & Within distance $\mathtt{n}$ \\ 
    \end{tabular}
\end{table}
 
\noindent \textbf{\emph{Road Variables.}}
Road variables, summarised in Table~\ref{tab:road_variables}, capture properties of the road the AV is currently driving on, e.g.~whether or not honking is allowed, the street light is on, or whether a junction is within $n$ units of distance ahead of the AV.
Most of the variables are self-explanatory, but three of them are based on special types.

First, $\mathtt{currentLane}$ returns the lane that the AV is currently on. This is an object $\mathtt{lane}$ of type $\mathtt{Lane}$, containing information such as the number of the current lane ($\mathtt{lane.number}$), the side of the road that the lane is on ($\mathtt{lane.side}$), and the allowed direction of travel ($\mathtt{lane.direction}$).
This can include values such as $\mathtt{left}$, $\mathtt{right}$, $\mathtt{UTurn}$, and various other combinations (e.g.~$\mathtt{forwardOrRight}$).

Second, $\mathtt{speedLimit}$ returns an object $\mathtt{limit}$ of type $\mathtt{SpeedLimit}$.
This contains information such as the lower ($\mathtt{limit.lowerLimit}$) and upper ($\mathtt{limit.upperLimit}$) speed limits of the road.
If a road does not have speed limits, the default values of these attributes will respectively be $-\infty$ and $+\infty$.

Finally, $\mathtt{specialLocationAhead(n)}$ returns an object $\mathtt{location}$ of type $\mathtt{SpecialLocation}$.
This contains attribute $\mathtt{location.type}$, which can have one of the following values: $\mathtt{Railway\_J}$, $\mathtt{Bridge}$, $\mathtt{SharpTurn}$, $\mathtt{SteepSlope}$, $\mathtt{Tunnel}$, $\mathtt{ArchBridge}$, $\mathtt{Slope}$, $\mathtt{Flooded}$, \\$\mathtt{OnewayRoad}$, $\mathtt{Roundabout}$, or $\mathtt{None}$ (i.e.~none of the above).\\

\begin{table}[!t]
	\centering
	\caption{Signal variables in \coolname}
	\label{tab:traffic_signal_variables}
	\footnotesize
    \begin{tabular}{c|c|p{1.5in}}
         Variable & Type & Remarks  \\
         \hline
         $\mathtt{stopSignAhead(n)}$ & Bool & Within distance $\mathtt{n}$ \\ 
         $\mathtt{noUTurnSignAhead(n)}$ & Bool & Within distance $\mathtt{n}$ \\ 
         $\mathtt{signalAhead}$ & Enum & $\mathtt{Common}$, $\mathtt{Arrow}$, or $\mathtt{None}$ \\ 
         $\mathtt{trafficLightAhead}$ & Signal & -- \\ 
    \end{tabular}
\end{table}

\noindent \textbf{\emph{Signal Variables.}}
Signal variables, summarised in Table~\ref{tab:traffic_signal_variables}, allow for the specification of laws involving traffic lights and various signs (e.g.~stop signs) at the junction an AV is approaching.
Two of the variables are self-explanatory Bool types, but the other two consist of richer data. First, $\mathtt{signalAhead}$ is of enumerated type. The value $\mathtt{Common}$ indicates a traffic light in which all lights are circles, whereas the value $\mathtt{Arrow}$ indicates one in which (some of) the lights are arrows.
$\mathtt{None}$ indicates that there is no traffic light ahead. Second, $\mathtt{trafficLightAhead}$ returns an object $\mathtt{signal}$ of type $\mathtt{Signal}$.
This consists of information such as the current colour of the traffic light ($\mathtt{signal.color}$), which can be $\mathtt{yellow}$, $\mathtt{green}$, $\mathtt{red}$, or $\mathtt{black}$; whether or not the light is blinking ($\mathtt{signal.isBlinking}$); and $\mathtt{signal.arrow}$, which returns an object $\mathtt{arrow}$ of type $\mathtt{SignalArrow}$ that contains similar information to $\mathtt{Signal}$ objects but also the direction of the arrow. \\

\begin{table}[!t]
	\centering
	\caption{Traffic variables in \coolname}
	\label{tab:traffic_variables}
	\footnotesize
    \begin{tabular}{c|c|p{1.45in}}
         Variable & Type & Remarks  \\
         \hline
         $\mathtt{PriorityNPCAhead(l/r)}$ & Bool & Vehicle with right of way \\ 
         $\mathtt{PriorityPedsAhead(l/r)}$ & Bool & Pedestrian with right of way \\
         $\mathtt{NPCAhead}$ & NPC & -- \\
         $\mathtt{NPCBack}$ & NPC & -- \\
         $\mathtt{NPCLeft}$ & NPC & -- \\
         $\mathtt{NPCRight}$ & NPC & -- \\
         $\mathtt{nearestNPC}$ & NPC & -- \\
         $\mathtt{NPCOpposite}$ & NPC & -- \\
         $\mathtt{isTrafficJam}$ & Bool & -- \\
    \end{tabular}
\end{table}

\noindent \textbf{\emph{Traffic Variables.}}
Traffic variables, summarised in Table~\ref{tab:traffic_variables}, are associated with other vehicles (NPCs) sharing the road with the AV, as well as any pedestrians crossing it. $\mathtt{PriorityNPCAhead(l/r)}$ indicates if there is an NPC vehicle ahead with priority over the ego vehicle, e.g.~due to priority at junctions, or due to the NPC being an ambulance.
The variable $\mathtt{PriorityPedsAhead(l/r)}$ indicates that there is a pedestrian with priority right of way ahead: pedestrians on a crosswalk, for example, have higher priority and are considered to be `ahead' if they are within five metres of the ego vehicle.
In both cases, the ego vehicle must not hinder the movement of the priority NPC/pedestrian when turning.
Note that for the customisation of traffic laws across different countries, we can use $\mathtt{l/r}$ to represent the different sides of the driver position: $\mathtt{PriorityNPCAhead(l)}$ and $\mathtt{PriorityPedsAhead(l)}$ if the driver position of the country/region is at the left side, and $\mathtt{r}$ if it is at the right. By default, the driver position will be treated as left for both variables.

The variables $\mathtt{NPCAhead}$, $\mathtt{NPCBack}$, $\mathtt{NPCALeft}$, $\mathtt{NPCRight}$,\\ $\mathtt{nearestNPC}$, and $\mathtt{NPCOpposite}$ respectively represent the NPC vehicle in front of the ego vehicle, the one behind, the one on the left, the one on the right, the one that is closest, and the one that is opposite.
These variables return objects $\mathtt{npc}$ of type of $\mathtt{NPC}$, which contain information such as the speed of the NPC vehicle ($\mathtt{npc.speed}$), the direction of the NPC vehicle ($\mathtt{npc.direction}$), the type of the NPC vehicle ($\mathtt{npc.type}$), as well as $\mathtt{npc(n)}$, which is true if the NPC is within $n$ units of distance from the ego vehicle. Note that the value of $\mathtt{npc.type}$ can have one of the following values: bus, car, priorityVehicle, or None.\\

\begin{table}[!t]
	\centering
	\caption{Map variables in \coolname}
	\label{tab:map_variables}
	\footnotesize
    \begin{tabular}{c|c|p{1.5in}}
         Variable & Type & Remarks  \\
         \hline
         $\mathtt{weather}$ & Weather & Current weather conditions \\
         $\mathtt{time}$ & Time & Current day/time \\
    \end{tabular}
\end{table}

\noindent \textbf{\emph{Map Variables.}}
The map variables, shown in Table~\ref{tab:map_variables}, are used to specify traffic laws related to environment conditions, e.g.~the weather or time of day. The variable $\mathtt{weather}$ returns an object $\mathtt{w}$ of type $\mathtt{Weather}$, consisting of information such as the degree of rain ($\mathtt{w.rain}$, valued from 0 to 1), degree of fog ($\mathtt{w.fog}$), or degree of snow ($\mathtt{w.snow}$), and the current visibility in metres. \\

\noindent \emph{\textbf{Semantics.} } Our specifications are interpreted over \emph{execution traces} from the ADS.
An execution trace $\pi$ is a sequence of scenes, denoted as {\small $\pi=\langle \theta_0, \theta_1, \ldots, \theta_n \rangle$}. 
A scene $\theta$ is a tuple of the form {\small $\theta=( f_0, f_1, \ldots, f_x )$} where $f_i$ is the valuation of all of the above-mentioned variables.

Given a trace $\pi$, we write $\pi \vDash \Phi$ (resp.~$\pi \nvDash \Phi$) to denote that $\Phi$ evaluates to be true (resp.~false) given trace $\pi$.
We use the standard definition of $\vDash$ for STL formulas (see e.g.~\cite{maler2004monitoring}).

\subsection{Case Study: Modelling China's Traffic Laws}
\label{sec:case_study}

Extending the illustrative example from Section~\ref{sec:overview_and_background}, as our main case study, we examined all of the traffic laws in the \emph{Regulations for the Implementation of the Road Traffic Safety Law of the People's Republic of China}~\cite{China_traffic_law}.
We labelled each rule with the following flags: \emph{relevant}, if the rule constrains an AV's behaviour in some way; \emph{describable}, if it can be specified using \coolname; and \emph{testable}, if the rule can potentially be tested in existing simulators (e.g.~LGSVL).
A summary of the labels for China's traffic laws is given in Table~\ref{tab:chinese_traffic_rule_summary}, and fully translated formulas for the describable laws can be found on our website~\cite{ourweb}.
(Our supplementary material~\cite{ourweb} also includes detailed translations of Singapore's traffic laws, demonstrating the generality of the language.)

While all laws are relevant, some of them are not describable in \coolname.
A typical example is Article \#65, which requires drivers to ``obey the instructions'' of ferry management personnel (this is too vague for our language to describe).
A number of rules are not testable due to lack of support in the underlying simulator.
For example, Article \#43 regulates the behaviour of vehicles when crossing a railway track.
While our language and fuzzer can be extended to cover such situations, railway tracks are not yet supported in the current maps of the simulator.
Additional details are presented later, in Section~\ref{sec:evaluation}, as part of our evaluation.

We present two of the translated traffic laws to highlight how the language is used. \\

\noindent \emph{\textbf{Article \#42: Yellow Lights.}}
Article \#42 stipulates that when a vehicle passes over a junction with flashing yellow light, it needs to ensure safety when passing through:

\emph{``The flashing warning signal light is a yellow light that continues to flash, reminding vehicles and pedestrians to pay attention when passing through, and pass after confirming safety.''}

\begin{table}[]
\footnotesize
\centering
\caption{Summary of Chinese traffic rules}
\label{tab:chinese_traffic_rule_summary}
\begin{tabular}{ccccc}
\hline
Category & Metric 
& Relevant  & Describable  & Testable  \\\hline
\multirow{2}{*}{General} 
& Count     & 8     &  8         & \textbf{3} \\ 
& Percent   & 100\% & 100\%   & \textbf{37.5\%}  \\ \hline
\multirow{2}{*}{Vehicles} 
& Count     & 40    & 37         & \textbf{21}  \\ 
& Percent   & 100\% & 92.5\%  & \textbf{52.5\%} \\ \hline
\multirow{2}{*}{Highway} 
& Count     & 12    &  12         & \textbf{0} \\ 
& Percent   & 100\% & 100\%     & \textbf{0\%} \\ \hline
\multirow{2}{*}{Others} 
& Count   & 0  & 0  & 0               \\ 
& Percent  & -  & -  & -              \\ \hline
\end{tabular}
\end{table}

This is an example of an ambiguous (or vague) traffic law that requires a specific formalisation in \coolname.
In particular, we can specify it as follows:

\begin{lstlisting}
proposition1 = trafficLightAhead.color==yellow & trafficLightAhead.blink &  (stoplineAhead(10) | junctionAhead(10));
proposition2 = F[0,2]((speed<20) U ~NearestNPC(15));
law42 = G(proposition1 -> proposition2);
\end{lstlisting}
The above translation expresses that when a flashing warning signal is $v_1$ meters ahead, the ego vehicle should move at a speed of less than $v_2\ km/h$ until there are no other vehicles within $v_3$ meters in the coming $v_4$ time steps. 
The value of variables $v_1, v_2, v_3, v_4$ can be customised by the user: here, we instantiated them with the values $10$, $20$, $15$, and $2$.
 
Note that the user can also customise these traffic laws in their own way, for instance, users can add constraints regarding pedestrians ahead by using the signal variable $\mathtt{PriorityPedsAhead}$ in the formula. \\

\noindent \emph{\textbf{Article \#52: Priority.}}
Article \#52 stipulates priority issues regarding vehicles, in particular, that the vehicle should give way to the vehicles with higher priority:

\emph{
``When a motor vehicle passes through an intersection that is not controlled by traffic lights or commanded by traffic police, in addition to complying with the provisions of Article \#51 (2) and (3), it shall also comply with the following provisions:
\begin{enumerate}
    \item[(1)] If there are traffic signs and markings, let the party with priority go first;
    \item[(2)] If there is no traffic sign or marking control, stop and look at the intersection before entering the intersection and let the traffic on the right road go first;
    \item[(3)] Turning motor vehicles let vehicles going straight go first;
    \item[(4)] A right-turning motor vehicle driving in the opposite direction should let the left-turning vehicle go first.''
\end{enumerate}
}

This is an example of how our driver-oriented specification approach can lead to a simpler specification than an equivalent globally specified property (as in AVUnit~\cite{AVUnit2021}).
In \coolname, we can describe sub-rules 2--4 of Article \#52 in the following way:
\begin{lstlisting}
prop1 = signalAhead==None & junctionAhead(0.5);
prop2 = F[0,2]((speed<0.5)U(~PriorityNPCAhead));
law52 = G(prop1 -> prop2);
\end{lstlisting}
The above translation expresses that when there is an intersection without traffic lights ahead, the ego vehicle is expected to stop until there is no vehicle with higher priority ahead. In order to simplify the specification of this traffic law, we utilise the signal variable $\mathtt{PriorityNPCAhead}$ to check whether there is a priority vehicle ahead. The design principle behind $\mathtt{PriorityNPCAhead}$ can be found in Section~\ref{sec:syntax_semantics}: ultimately, this signal variable allows a direct translation of the traffic law. \\

%% file: Fuzzing.tex
\section{The LawBreaker Fuzzing Engine}
\label{sec:Fuzzing}
In the real world, traffic laws can be violated in multiple different ways.
For instance, as shown in Section~\ref{sec:overview_and_background}, there are two ways to violate Article \#38's yellow light sub-rule: failing to move when the ego vehicle has already reached the stop line, and failing to stop when the ego vehicle is at a safe stopping distance before it.
Both of these violations are interesting, but for very different reasons: the first leads to an AV that is driving less efficiently than it (legally) might, whereas the second is quite dangerous and could increase the likelihood of a traffic accident.
In general, we need a fuzzing approach that can find as many different ways of violating a specification as possible, as multiple test cases will help to pinpoint the key problems in the ADS.

In this section, we present a fuzzing algorithm based on the idea of: (1)~identifying the different possible ways of violating a law specification $\Phi$ in our language, then (2)~searching for concrete test cases that come `closer' (as measured by a quantitative semantics) to violating $\Phi$ in the different ways that were identified.

\subsection{Specification Violation Coverage}
\label{sec:sepcification_sub_violation_defination}
Given a law specification $\Phi$, we write $\Theta(\Phi)$ to denote a set of constraints that represents different ways in which $\Phi$ might be violated. 
Formally, $\Theta(\Phi)$ is a set of formulas which satisfies the following proposition:
\begin{proposition}
\label{prop:overall}
Let $\Phi$ be an STL formula and $\pi$ a trace. Then:
\[\forall \xi \in \Theta(\Phi).~\pi \vDash \xi \implies \pi \nvDash \Phi\]\hfill $\qed$
\end{proposition}

In \coolname, $\Theta(\Phi)$ is computed as follows: \\

\begin{math}
\begin{aligned}
 \Theta(\mu) = &~ \{\neg \mu\} , where\ \mu\ is\ a\ Boolean\ expression & \\
 \Theta(\theta_1 \land \theta_2) = &~ \Theta(\theta_1) \cup \Theta(\theta_1) \\
 \Theta(\theta_1 \lor \theta_2)= &~ \{x \land y ~|~ x \in \Theta(\theta_1) \land y \in \Theta(\theta_2)\} \\
 \Theta(\neg \theta_1) = &~ N(\theta_1) \\
 \Theta(\Box_\mathcal{I} \theta_1) = &~  \{\Diamond_\mathcal{I}\ \theta ~|~ \theta \in \Theta(\theta_1)\} \\
 \Theta(\Diamond_\mathcal{I} \theta_1) = &~ \{\Box_\mathcal{I}\ \theta ~|~ \theta \in \Theta(\theta_1)\} \\
\Theta(\theta_1 \mathcal{U_I} \theta_2) = &~ \{x~\mathcal{U_I}~y ~|~ x \in \Theta(\lnot\theta_1\lor\theta_2) \land y \in \Theta(\theta_1 \lor \theta_2)\} \\
  & \cup \{x \land y ~|~ x \in \Theta(\theta_1) \land y \in \Theta(\theta_2)\} \\
\Theta(\bigcirc \theta_1) = &~ \{\bigcirc x ~|~ x \in \Theta(\theta_1) \}
\end{aligned}
\end{math}\\

\noindent where $N(\Phi)$ represents different ways in which $\Phi$ might be satisfied, i.e., $N(\Phi)$ is a set of formulas satisfying the following condition: \[\forall \xi \in N(\Phi).~\pi \vDash \xi \implies \pi \vDash \Phi\]

\noindent It is systematically computed as follows: \\

\begin{math}
\begin{aligned}
N(\mu) =&~ \{\mu\}, where\ \mu \ is\ a\ Boolean\ expression\\
N(\theta_1 \land \theta_2) =&~ \{x \land y ~|~ x \in N(\theta_1) \land y \in N(\theta_2)\} \\
N(\theta_1 \lor \theta_2) =&~ N(\theta_1) \cup N(\theta_2) \\
N(\neg \theta_1) =&~ \Theta(\theta_1) \\
N(\Box_\mathcal{I} \theta_1) =&~ \{\Box_\mathcal{I}\ \theta ~|~ \theta \in N(\theta_1)\} \\
N( \Diamond_\mathcal{I} \theta_1) =&~ \{\Diamond_\mathcal{I}\ \theta ~|~ \theta \in N(\theta_1)\} \\
N(\theta_1\ \mathcal{U_I}\ \theta_2) =&~ \{x \ \mathcal{U_I}\ y ~|~ x \in N(\theta_1) \land y \in N(\theta_2)\} \\
N(\bigcirc \theta_1) =&~ \{\bigcirc \theta ~|~ \theta \in N(\theta_1)\} \\
\end{aligned}
\end{math} \\

\noindent Proposition~\ref{prop:overall} can be proven by structural induction. The detailed proof can be found in our supplementary materials~\cite{ourweb}. 

\begin{example}
The value of $\Theta(\Phi)$ is calculated recursively.
For example, given a specification $\Phi = \Box ((a \lor b) \to c)$, before the calculation of $\Theta$, we first pre-process $\Phi$ to get $\Phi' = \Box (\lnot(a \lor b) \lor c)$. The formula $\Phi$ is equivalent to $\Phi'$. 
Then, we can get $\Theta(\Phi)$ as follows:
\begin{enumerate}
    \item Calculation of the primitive elements:
    
    $\Theta(c) = \{\lnot c\}, N(a) = \{a\}, N(b) = \{b\}$
    \item Given $N(a) = \{a\}$ and $N(b) = \{b\}$: $N(a \lor b) = \{a, b\}$
    \item Given $N(a \lor b)$: $\Theta(\lnot (a \lor b))= \{a, b\}$
    \item Given $\Theta(\lnot (a \lor b))$ and $\Theta(c)$:

    $\Theta(\lnot (a \lor b) \lor c) = \{a \land \lnot c, b\land \lnot c\}$
    \item Given $\Theta(\lnot (a \lor b) \lor c)$:

    $\Theta(\Box (\lnot (a \lor b) \lor c)) = \{\Diamond (a \land \lnot c), \Diamond (b\land \lnot c ) \} $

    \item The final result is $\Theta(\Phi) = \{\Diamond (a \land \lnot c), \Diamond (b\land \lnot c ) \} $.
\end{enumerate}

\noindent i.e.~we can `cover' the different ways of violating the original specification $\Phi$ by finding traces that satisfy $\Diamond (a \land \lnot c)$ and $\Diamond (b\land \lnot c )$.
\end{example}

\subsection{Quantitative Semantics}
\label{sec:Quantitative_Semantics}
Given a formula $\Phi$, the overall idea of our fuzzing algorithm is to systematically generate test cases to violate each STL formula $\varphi \in \Theta(\Phi)$, if feasible. We thus first adopt a quantitative semantics for our specification language, which allows us to iteratively generate test cases that come `closer' to violating a given formula.

The quantitative semantics is adopted from~\cite{maler2004monitoring, deshmukh2017robust,nivckovic2020rtamt}, which, intuitively speaking, defines the semantics of a formula $\varphi$ with respect to a trace $\pi$ in the form of a \emph{robustness value}, i.e.~a number representing how far $\varphi$ is from being satisfied by $\pi$.
Our algorithm then attempts to maximise this number so as to generate a violation.

\begin{definition}[Quantitative Semantics]\label{def:Quantitative_Semantics}
Given a trace $\pi$ and a formula $\varphi$, the quantitative semantics is defined as the robustness degree $\rho(\varphi, \pi,t)$ where $t$ is the time step:
{\small\begin{align*}
\rho(\mu,\pi,t) & = f(\pi)(t), \\
\rho(\neg\varphi,\pi,t) & = -\rho(\varphi,\pi,t), \\
\rho(\varphi_1 \land \varphi_2,\pi,t) & = \min\{\rho(\varphi_1,\pi,t),\rho(\varphi_2,\pi,t)\}, \\
\rho(\varphi_1 \lor \varphi_2,\pi,t) & = \max\{\rho(\varphi_1,\pi,t),\rho(\varphi_2,\pi,t)\}, \\
\rho(\varphi_1 \;\mathcal{U_I}\; \varphi_2,\pi,t) & = \sup_{t'\in t+I} \min \{\rho(\varphi_2,\pi,t'), \inf_{t'' \in [t,t']} \rho(\varphi_1,\pi,t'')\}, \\
\rho(\Diamond_\mathcal{I}\varphi,\pi,t) & = \sup_{t'\in t+\mathcal{I}}\rho(\varphi,\pi,t'), \\
\rho(\Box_\mathcal{I}\varphi,\pi,t) & = \inf_{t'\in t+\mathcal{I}}\rho(\varphi,\pi,t'),\\
\rho(\bigcirc \varphi,\pi,t) & = \rho(\varphi,\pi,t+1).
\end{align*}}
\end{definition}

If $\rho(\varphi,\pi,t)$ is equal to or greater than $0$, we successfully found a way to violate $\Phi$. Note that $\rho(\varphi,\pi)$ is treated as $\rho(\varphi,\pi,0)$.

\begin{example}
Given a traffic law {\small$\Phi = \Box (speed < 80)$} which means the speed of the ego vehicle should always be less than $80km/h$, suppose the $speed$ over a trace $\pi$ is {\small$\{(t, s): (t_0, 0)$, $(t_1, 0)$, $(t_2, 0.5)$, $(t_3, 1.8)$, $(t_4, 4.5)$, $(t_5, 7.9)$, $(t_6, 10.9)$, $\ldots$ , $(t_{39}, 85)$, $\ldots\}$} where the maximum value of $speed$ is $85km/h$ at time step $t_{39}$, and the relevant specification {\small$\varphi \in \Theta(\Phi)$} is {\small$\varphi = \Diamond (speed > 80)$}, then we have {\small$\rho(\varphi, \pi) = \rho(\varphi, \pi, 0) = max_{t' \in [0, |\pi|]} ( speed(t') -80 ) = 5 > 0$}. It means the specification $\varphi$ is satisfied by $\pi$ which leads to a violation of the traffic law $\Phi$.
\end{example}

\subsection{Genetic Encoding for Scenarios} 
\label{sec:background_of_genetic_alg}
Our fuzzing algorithm is based on a genetic algorithm~(GA)~\cite{mirjalili2019genetic}, and requires an appropriate genetic encoding for the targeted scenario description language, as well as a customisation of the crossover and mutation operators.

In this work, we adopt the AVUnit~\cite{AVUnit2021} scenario description language, which describes a test case in terms of the ego vehicle, NPC vehicles, pedestrians, static obstacles, and the environment (e.g.~weather).
We encode scenarios in terms of their operable parameters.
For the ego vehicle, the operable parameter is its starting point.
For NPCs, all parameters are operable except the speed at their destinations (which is always zero).
Finally, for static obstacles and the environment, all parameters are operable.
Note that in valid encodings, vehicles must always be positioned within a lane or junction area, pedestrians may move around empty regions of the map, and obstacles can be placed anywhere (e.g.~a basketball on the road, or even a meteorite on the crosswalk).

For parameters with continuous values (e.g.~position, speed, and weather), we apply Gaussian mutation.
We also apply the clipping function to avoid invalid values. 

For each pair of test cases, the genetic crossover operation can only be done in the same category (e.g.~position, NPC type).
In theory, we can perform crossover for each category, but to avoid generating infeasible scenarios, we do not perform crossover on the chromosomes encoding vehicle and pedestrian positions.

Note that since the crossover and mutation operations are always limited to the valid space of the operable parameters, the newly generated test cases are always valid.

\subsection{Fuzzing Algorithm}
\label{sec:fuzzing_algorithm}

\begin{algorithm}[t]
\caption{\coolname Fuzzing Algorithm}\label{alg:fuzzing_for_law_breaking}
\small
\KwIn{$\Phi$, $n$ (population size), and $M$ (maximal generations)}
\KwOut{A test suite  $\Gamma$}
Let $\Theta_r = \Theta(\Phi)$ be the set of uncovered formulas in $\Theta(\Phi)$\;
Let $Seeds_r$ be a mapping from $\Theta_r$ to tests, initially empty\;
Let $Robust_r$ be a mapping from $\Theta_r$ to robustness, initially $-\infty$\;
Let $G=\{s_1,\cdots,s_n\}$ be a set of randomly generated tests\;
Let $\Gamma$ be an empty set\;
\While{$\Theta_r$ is not empty and not timeout}
{
    \For{each $s_i$ in $G$ }{
        Execute $s_i$ via simulation and obtain trace $\pi_i$\;
        \For{each $\xi$ in $\Theta_r$ }{
            Compute the robustness $\rho(\xi, \pi_i)$\; 
            \If{$\rho(\xi, \pi_i) \geq 0$}{
                Remove $\xi$ from $\Theta_r$; Add $s_i$ into $\Gamma$\;
            }
            \ElseIf{$\rho(\xi, \pi_i) > Robust_r(\xi)$} {
                    Set $Seeds_r(\xi)$ to be $s_i$\;
                    Set $Robust_r(\xi)$ to be $\rho(\xi, \pi_i, t)$\;
            }
        }
    }
    Set $G$ to be a set of new test cases generated based on $Seeds_r$ through selection, mutation and crossover\;
}
\Return $\Gamma$
\end{algorithm}

We are now ready to present the overall fuzzing algorithm of \coolname, which directly utilises $\Theta(\Phi)$ (i.e.~the different possible ways to violate $\Phi$) and $\rho(\xi,\pi)$ (i.e.~how `close' we are to violating the formula $\xi$). Note that $\xi \in \Theta(\Phi)$.

Our fuzzing approach is detailed in Algorithm~\ref{alg:fuzzing_for_law_breaking}. First, we generate some initial test cases randomly then initialise $Seeds_r$ as empty and $Robust_r$ as $-\infty$. 
Note that $Seeds_r$ and $Robust_r$ are mappings from the remaining formulas in $\Theta_r$ to the corresponding test cases and robustness scores. 
For every test case in a generation, we execute it to obtain trace $\pi$ and compute the robustness $\rho(\xi, \pi)$\ for all the uncovered $\xi \in \Theta_r$. We remove $\xi$ once it is satisfied, i.e.~$\rho(\xi, \pi_i) \geq 0$, and add the corresponding scenario $s_i$ to the output set $\Gamma$. If some $\xi$ is not satisfied, we update $Seeds_r(\xi)$ and $Robust_r(\xi)$ when the current trace is closer to the satisfaction of $\xi$, i.e.~$\rho(\xi, \pi_i) > Robust_r(\xi)$. 

After executing and processing all of the test cases of a population, we generate the population for the next generation based on $Seeds_r$ and $Robust_r$.
Since the size of $Seeds_r$ is equal to the size of $\Theta_r$, it may be larger or smaller than the population size, and thus we select parents as follows.
We first sort $Seeds_r$ according to $Robust_r$ in descending order. 
Note that the greater the robustness value is, the more likely the test case leads to a new violation of the traffic law $\Phi$.
To add some uncertainty, we first choose an individual from the first half of the population,
and then we select an individual from the overall population by random sampling. From these two selected individuals, the individual with the higher score of robustness is chosen. Note that the higher the score,  the closer the test case is to the targeted violation.
We repeat this selection process until a given number of the parent population is selected. (This number can be chosen by the user.)

With the parents set selected, we apply crossover and mutation as defined in Section~\ref{sec:background_of_genetic_alg} to obtain the next generation. 
This process is repeated until all the elements in set $\Theta(\Phi)$ are covered or the maximal number of generations has been reached. Note that the maximal number of generations $M$ is defined by the user.

%% file: Evaluation.tex
\section{Implementation and Evaluation}
\label{sec:experiments}

In this section, we present our implementation and evaluation of \coolname based on an existing popular simulation framework.

\subsection{Implementation}
Implementing \coolname for a given ADS and simulator requires the completion of the following three steps: (1)~construction of a \emph{bridge} for collecting messages from the ADS and spawning scenarios in the simulator; (2)~implementation of a \emph{library} for converting those messages into signals for the traffic law language in Section~\ref{sec:Spec}; and (3)~implementation of the \emph{fuzzing engine} in Section~\ref{sec:Fuzzing}.
Our source code of all three components for Apollo+LGSVL is available online~\cite{ourweb}.

In this work, we implemented a bridge for the Apollo 5.0 and Apollo 6.0 ADSs.
Our bridge retrieves messages from the ADS that include the driving status of the ego vehicle (e.g.~speed, position, high beam) and the environment (e.g.~status of NPC vehicles and nearby traffic signals) at different time steps.
Currently, the perception part of both Apollo versions is still under development, so we follow the recommendation of the vendor and use the ground truth as the input of the $\mathtt{PerceptionObstacle}$ module.
Our bridge allows for all the operable parameters mentioned in Section~\ref{sec:background_of_genetic_alg} (e.g.~trajectories of NPC vehciles and pedestrians) to be translated to API calls in the simulator, thus spawning scenarios based on our genetic encoding.

With the bridge retrieving the original messages from the ADS, we then use our library to translate these messages into the signal variables described in Section~\ref{sec:Spec}.
Some signal variables are quite intuitive.
For instance, the signal variables $\mathtt{speed, acc, brake}$ are the speed, acceleration, and brake percentages of the ego vehicle, and we obtain these through a simple analysis of ADS messages from different time steps.
However, some signal variables require more complex processing.
For example, it takes a few steps to calculate the value of signal $\mathtt{PriorityPedsAhead}$ at time step $\mathtt{t}$.
First, we calculate the `area ahead' with respect to the current position and direction of the ego vehicle.
Then, we check whether there is a pedestrian within that area based on the positions of pedestrians.
If so, we check whether the pedestrian is likely to cross based on their distance to a crosswalk.
If both are satisfied, then the value of the signal $\mathtt{PriorityPedsAhead}$ at that time step is set to be true, otherwise false.

Finally, we implemented the fuzzing engine as described in Section~\ref{sec:fuzzing_algorithm}.
Our implementation supports six kinds of mutations, i.e., mutation of position, speed, time, weather, NPC vehicle type, and pedestrian type.
Furthermore, we embedded the tool \emph{RTAMT}~\cite{nivckovic2020rtamt} to compute the robustness of the specifications with respect to the trace obtained from the bridge.

\subsection{Evaluation}
\label{sec:evaluation}
In the following, we conduct multiple experiments to answer our key Research Questions~(RQs). Since \coolname is designed for the description and evaluation of traffic laws across different countries, we evaluate it from three aspects: versatility, effectiveness, and efficiency, which correspond to our three RQs. \\

\begin{table}[]
\footnotesize
\centering
\caption{Testing AVs, where $\times$, $\Delta$, and $\surd$ means no support, limited support and full support respectively}
\label{tab:universitality_of_language}
\scalebox{0.9}{
\begin{tabular}{|>{\centering}m{0.08\linewidth}|c|>{\centering}m{0.03\linewidth}|>{\centering}m{0.03\linewidth}|>{\centering}m{0.03\linewidth}|>{\centering}m{0.03\linewidth}|>{\centering}m{0.03\linewidth}|>{\centering}m{0.03\linewidth}|>{\centering}m{0.03\linewidth}|>{\centering}m{0.03\linewidth}|m{0.22\linewidth}|}
\hline
\multicolumn{2}{|c|}{  \multirow{2}{*}{\textbf{Traffic Laws}}}  & FIH & FMI & FTL & FA & \multicolumn{2}{c|}{\textbf{AVUnit}} & \multicolumn{3}{c|}{ \textbf{LawBreaker} } \\ \cline{3-11}
\multicolumn{2}{|c|}{} & D & D & D & D & D & T & D & T & Why $\lnot D \lor \lnot T$ \\ \hline
\multicolumn{2}{|c|}{Law38 sub1-3}   & $\times$ & $\times$    & $\times$  & $\times$  & $\Delta$  & $\Delta$ & $\surd$  & $\surd$ & -\\ \hline  
\multicolumn{2}{|c|}{Law40-43}    & $\times$ & $\times$   & $\times$  & $\times$ & $\times$  & $\times$ & $\surd$  & $\times$ & Lack Map Support\\ \hline 
\multicolumn{2}{|c|}{Law44}   & $\times$ & $ \Delta$ & $\Delta$  & $\times$ & $\times$  & $\times$ & $\surd$  & $\surd$ & -\\ \hline       
\multirow{2}{*}{Law45}  & sub1   & $\times$ & $\surd$ & $\surd$ & $\Delta$ & $\surd$  & $\surd$ & $\surd$  & $\surd$ & -\\ \cline{2-10} 
                            & sub2  & $\times$ & $\surd$  & $\surd$ & $\Delta$ & $\surd$  & $\surd$ & $\surd$  & $\surd$ & -\\ \hline
\multirow{4}{*}{Law46}  & sub1 & $\times$ & $\times$   & $\Delta$  & $\times$ & $\times$  & $\times$ & $\surd$  & $\times$ & Lack Map Support\\ \cline{2-10} 
                            & sub2  & $\times$ & $\times$  & $\times$  & $\times$ & $\times$  & $\times$ & $\surd$  & $\Delta$  & Lack Map Support\\ \cline{2-11}
                            & sub3  & $\times$ & $\times$  & $\times$  & $\times$ & $\times$  & $\times$ & $\surd$  & $\surd$  & -\\ \cline{2-11}
                            & sub4  & $\times$ & $\times$ & $\times$  & $\times$ & $\times$  & $\times$ & $\surd$  & $\times$ & Lack Map Support\\ \hline
\multicolumn{2}{|c|}{Law47}    & $\surd$ & $\Delta$   & $\Delta$  & $\times$ & $\times$  & $\times$ & $\surd$  & $\surd$ & -\\ \hline  
\multicolumn{2}{|c|}{Law48 sub1-2}  & $\times$ & $\Delta$   & $\Delta$  & $\Delta$ & $\times$  & $\times$ & $\surd$  & $\times$ & Lack Map Support\\ \hline 
\multicolumn{2}{|c|}{Law48 sub3-4}    & $\times$ & $\times$   & $\times$  & $\times$ & $\times$  & $\times$ & $\times$  & $\times$ & Vague\\ \hline 
\multirow{1}{*}{Law48}  & sub5   & $\times$ & $\times$   & $\times$  & $\times$ & $\times$  & $\times$ & $\surd$  & $\times$ & Lack Map Support\\ \hline 

\multicolumn{2}{|c|}{Law49}   & $\times$ & $\times$    & $\times$  & $\times$ & $\times$  & $\times$ & $\surd$  & $\times$ & Lack Map Support\\ \hline  
\multicolumn{2}{|c|}{Law50}   & $\times$ & $\times$   & $\Delta$  & $\times$ & $\times$  & $\times$ & $\surd$  & $\surd$ & -\\ \hline  
\multicolumn{2}{|c|}{Law51 sub1-2}  & $\times$ & $\times$    & $\times$  & $\times$ & $\times$  & $\times$ & $\surd$  & $\times$ & Lack Map Support\\ \hline  
\multirow{1}{*}{Law51}    & sub3   & $\times$ & $\times$    & $\times$  & $\times$ & $\times$  & $\times$ & $\surd$  & $\surd$ & -\\ \hline  
\multicolumn{2}{|c|}{Law51 sub4-7}  & $\times$ & $\times$     & $\times$  & $\times$ & $\Delta$  & $\Delta$ & $\surd$  & $\surd$ & -\\ \hline  
\multirow{1}{*}{Law52}    & sub1   & $\times$ & $\times$  & $\times$  & $\times$ & $\times$  & $\times$ & $\surd$  & $\times$ & Lack Map Support\\ \hline  
\multicolumn{2}{|c|}{Law52 sub2-4}   & $\times$ & $\Delta$  & $\times$  & $\times$ & $\Delta$  & $\Delta$ & $\surd$  & $\surd$ & -\\ \hline
\multicolumn{2}{|c|}{Law53}    & $\times$ & $\times$  & $\surd$ & $\times$ & $\Delta$  & $\Delta$ & $\surd$  & $\surd$ & -\\ \hline    
\multicolumn{2}{|c|}{Law57 sub1-2}  & $\times$ & $\times$  & $\times$  & $\times$ & $\times$  & $\times$ & $\surd$  & $\surd$ & -\\ \hline 
\multicolumn{2}{|c|}{Law58}   & $\times$ & $\times$  & $\times$  & $\times$ & $\times$  & $\times$ & $\surd$  & $\surd$ & -\\ \hline  
\multicolumn{2}{|c|}{Law59}    & $\times$ & $\times$  & $\times$  & $\times$ & $\times$  & $\times$ & $\surd$  & $\surd$ & -\\ \hline 
\multirow{1}{*}{Law62}  & sub1  & $\times$ & $\times$  & $\times$  & $\times$ & $\times$  & $\times$ & $\surd$  & $\times$ & Lack Sensors \\ \hline
\multirow{1}{*}{Law62}  & sub4  & $\times$ & $\times$  & $\times$  & $\times$ & $\times$  & $\times$ & $\surd$  & $\times$ & Lack Map Support\\ \hline
\multirow{1}{*}{Law62}  & sub8   & $\times$ & $\times$  & $\times$  & $\times$ & $\Delta$  & $\times$ & $\surd$  & $\times$ & Lack Map Support\\ \hline
\multicolumn{2}{|c|}{Law63 sub1-3} & $\times$ & $\times$  & $\times$  & $\times$ & $\times$  & $\times$ & $\surd$  & $\times$ & Lack Map Support\\ \hline 
\multicolumn{2}{|c|}{Law64}   & $\times$ & $\times$  & $\times$  & $\times$ & $\times$  & $\times$ & $\surd$  & $\times$ & Lack Map Support\\ \hline 
\multicolumn{2}{|c|}{Law65}   & $\times$ & $\times$  & $\times$  & $\times$ & $\times$  & $\times$ & $\times$  & $\times$ & Vague\\ \hline 
\multicolumn{2}{|c|}{Law78}   & $\times$ & $\Delta$  & $\Delta$  & $\Delta$ & $\surd$  & $\times$ & $\surd$  & $\times$ & Lack Map Support\\ \hline 
\multicolumn{2}{|c|}{Law 79-82}    & $\times$ & $\Delta$  & $\Delta$  & $\times$ & $\times$  & $\times$ & $\surd$  & $\times$ & Lack Map Support\\ \hline  
\multicolumn{2}{|c|}{Law84}     & $\times$ & $\times$  & $\times$  & $\times$ & $\Delta$  & $\times$ & $\surd$  & $\times$ & Lack Map Support\\ \hline 


\end{tabular}
}
\end{table} 

\noindent \emph{\textbf {RQ1: Can we test AVs against traffic laws using LawBreaker?}} To answer this question, we systematically examined all Chinese traffic laws related to AVs and determined whether or not they were describable and testable (as per Section~\ref{sec:case_study}) using \coolname.
In addition, we examine whether the same laws can be expressed using alternative specification approaches provided by other frameworks.

A number of existing works~\cite{AVUnit2021, rizaldi2015formalising, esterle2020formalizing, maierhofer2020formalization, rizaldi2017formalising} propose methods of evaluating AVs under different oracles, and we compare against the ones that are capable of specifying (at least some) traffic laws.
In particular, we compare against formalisations in Isabelle/HOL~(FIH)~\cite{rizaldi2017formalising}, for Machine Interpretability~(FMI)~\cite{esterle2020formalizing}, in Temporal Logic~(FTL)~\cite{maierhofer2020formalization}, for Accountability~(FA)~\cite{rizaldi2015formalising}, and finally, AVUnit's own test engine based on global specifications~\cite{AVUnit2021}.

Table~\ref{tab:universitality_of_language} presents the results of our evaluation.
Here, \textbf{D} means whether the framework allows description of the traffic law, and \textbf{T} means whether the framework can test the specific traffic law with the support of existing simulators. 
We observe that \coolname supports most of the relevant traffic laws and outperforms the existing works in this aspect, largely due to our driver-oriented language that allow specifications to be scenario-independent.
For example, for traffic lights, AVUnit's own test engine (based on global specifications) requires the user to be familiar with the map and to formulate specifications based on the IDs of every traffic light, the positions of every vehicle, and so on. Even more problematic is that users have to write a different specification for every specific scenario since the map is different. \coolname solves this problem by making use of the driver-oriented signal variable $\mathtt{trafficLightAhead}$, which makes it independent from the scenarios.

Focusing on \coolname, the last column of Table~\ref{tab:universitality_of_language} summarises the reason why \coolname is unable to describe or test a given Chinese traffic law.
There are several reasons.
First, it may be because the law is irrelevant to AVs (e.g.~they regulate the behaviour of pedestrians rather than the AVs); we do not list these in the table.
Second, it may be because the law is hard to evaluate or quantify.
For instance, sub4 of Law48~\cite{China_traffic_law}, which regulates priority on mountain roads, is rather vague.
Third, the law cannot be tested due to the limitation of existing maps.
For example, Law40 and Law41 regulate the behaviour of vehicles when facing traffic lights with arrow lights cannot be tested, since LGSVL does not support traffic lights with arrow lights (this is a limitation of the simulator, rather than \coolname).
Finally, some laws cannot be tested due to the lack of certain sensors.
For example, sub1 of Law62 requires that the doors and compartments must be closed when driving.
However, there is currently no sensor in the ADS for detecting the status of doors and compartments.

To summarise, \coolname is able to specify most of the relevant laws. The main reason why some laws cannot be tested is the limitation of the underlying simulators, i.e.~those laws can be tested in the future once sufficient map and sensor support are provided. \\

\noindent \emph{\textbf {RQ2: How effective is LawBreaker at generating violations of laws?}} To answer this question, we systematically apply our fuzzing algorithm to test all the testable laws. 
The results are summarised in Table~\ref{tab:coverage_of_failures}. Note that there are two different versions of an ADS driver being tested: \emph{Apollo5.0} and \emph{Apollo6.0}, which are the two latest versions of ADSs developed on the Apollo platform. The \emph{Violations} and \emph{Accidents} in the table denote whether the driver violates the traffic law and whether there are accidents due to the violations of the law.  Note that we mark $\surd$ for a traffic law $\Phi$ if and only if accidents happened in the trace $\pi$ and $\pi \nvDash \Phi$. Since ours is the first work which is capable of generating law-breaking test cases, we have no baseline to compare with in this experiment. 

As can be seen from the table, \coolname is able to trigger violations of most of the laws (sometimes in multiple ways). In summary, \coolname is able to find violations of 14 different Chinese traffic laws by Baidu Apollo. Among the test cases generated by our framework, 173 of them not only violate the laws but also \emph{cause accidents}. Furthermore, \emph{Apollo6.0} violates more traffic laws and results in more accidents than \emph{Apollo5.0}. 
While this is surprising, a close investigation shows that \emph{Apollo6.0} drives more aggressively than \emph{Apollo5.0} since the \emph{Apollo6.0} uses a deep learning model, while \emph{Apollo5.0} is completely controlled through a program.

In the following, we categorise the identified issues and present examples in each category. 
Video recordings of all the identified issues are available at~\cite{ourweb}. 
Note that we reran the corresponding test cases at least 3 times to ensure that all issues are reproducible. 

\begin{table}[]
\footnotesize
\centering
\caption{Violations of Chinese traffic laws}
\label{tab:coverage_of_failures}
\begin{tabular}{|>{\centering}m{0.1\linewidth}|>{\centering}m{0.05\linewidth}|>{\centering}m{0.06\linewidth}|>{\centering}m{0.06\linewidth}|>{\centering}m{0.06\linewidth}|>{\centering}m{0.06\linewidth}|c|}
\hline
\multicolumn{2}{|c|}{  \multirow{2}{*}{\textbf{Traffic Laws}}}  & \multicolumn{2}{c|}{ \textbf{Violations}} & \multicolumn{2}{c|}{\textbf{Accidents} } & \multirow{2}{*}{\textbf{Content}}\\ \cline{3-6}
\multicolumn{2}{|c|}{} & 5.0 & 6.0 & 5.0 & 6.0 & \multicolumn{1}{c|}{} \\ \hline
\multirow{3}{*}{Law38}  & sub1    & $\surd$  & $\surd$ & $\times$  & $\surd$  & green light\\ \cline{2-7} 
                            & sub2    & $\surd$  & $\surd$ & $\surd$  & $\surd$ & yellow light\\ \cline{2-7}
                            & sub3    & $\surd$  & $\surd$ & $\times$  & $\surd$ & red light\\ \hline
\multicolumn{2}{|c|}{Law44}       & $\surd$  & $\surd$ & $\surd$  & $\surd$ & lane change\\ \hline       
\multirow{2}{*}{Law45}  & sub1    & $\times$  & $\times$ & $\times$  & $\times$  & speed limit\\ \cline{2-7} 
                            & sub2    & $\times$  & $\times$ & $\times$  & $\times$ & speed limit\\ \hline
\multirow{2}{*}{Law46}  & sub2    & $\times$ & $\surd$ & $\times$ & $\times$  & speed limit\\ \cline{2-7} 
                            & sub3    & $\surd$  & $\surd$ & $\surd$  & $\surd$ & speed limit\\ \hline
\multicolumn{2}{|c|}{Law47}       & $\surd$  & $\surd$ & $\surd$  & $\surd$ & overtake\\ \hline  
\multicolumn{2}{|c|}{Law50}       & $\times$  & $\times$ & $\times$  & $\times$ & reverse\\ \hline  
\multirow{6}{*}{Law51}      & sub3    & $\times$  & $\surd$ & $\times$  & $\times$  & traffic light\\ \cline{2-7} 
                            & sub4    & $\surd$  & $\surd$ & $\surd$  & $\surd$  & traffic light\\ \cline{2-7} 
                            & sub5    & $\surd$  & $\surd$ & $\times$  & $\surd$  & traffic light\\ \cline{2-7} 
                            & sub6    & $\times$  & $\times$ & $\times$  & $\times$  & traffic light\\ \cline{2-7} 
                            & sub7    & $\times$  & $\times$ & $\times$  & $\times$ & traffic light\\ \hline
\multicolumn{2}{|c|}{Law52 sub2-4}       & $\times$  & $\times$ & $\times$  & $\times$ & priority\\ \hline 
\multicolumn{2}{|c|}{Law53}       & $\times$  & $\times$ & $\times$  & $\times$ & traffic jam\\ \hline           
\multirow{2}{*}{Law57}  & sub1    & $\surd$  & $\surd$ & $\times$  & $\times$  & left turn signal\\ \cline{2-7} 
                        & sub2    & $\surd$  & $\surd$ & $\times$  & $\times$ & right turn signal\\ \hline 
\multicolumn{2}{|c|}{Law58}       & $\surd$  & $\surd$ & $\surd$  & $\surd$ & warning signal\\ \hline  
\multicolumn{2}{|c|}{Law59}       & $\surd$  & $\surd$ & $\surd$ & $\surd$ & signals\\ \hline 
\multirow{1}{*}{Law62}  & sub8    & $\times$  & $\times$ & $\times$  & $\times$  & honk\\ \hline
\end{tabular}
\end{table} 

\emph{Dangerous behaviours.} The AV may break a law and result in dangerous behaviour. For instance, it might rush at a yellow light (violating Article \#38) and cause accidents. On the other hand, it might also hesitate at a yellow light and cross the junction at a red light, i.e. although the AV reaches the stop line when the traffic light turned from green to yellow and is expected to go across, it hesitates at the yellow light and crosses the intersection eventually at the red light. In another instance, the AV may fail to complete overtaking a large vehicle and cause accidents. In this situation, the AV is trying to overtake a large turning vehicle (e.g.~bus). But the ego vehicle wrongly estimates the distance to the large vehicle and accelerates, which causes collisions. The ego vehicle violates both Article \#44 and Article \#47 in this situation. Moreover, while the AV is expected to drive slowly with caution in heavy rain or fog, it ignores the weather condition and drives at a high speed (violating sub-rule3 of Article \#46). We remark that some of these behaviours do not result in accidents and thus would be missed by existing approaches~\cite{AVUnit2021, li2020av}. They are nonetheless behaviours that should be investigated and corrected.

 \emph{Inefficiency.}  The AV may break a law and result in inefficient driving. For instance, it might remain stationary while the traffic light is green. It might also hesitate at a yellow light---although it is safe to cross---until the traffic light turns red. Furthermore, it might fail to overtake a stationary vehicle ahead at an intersection. That is, there is a static NPC vehicle ahead and the traffic light turns to green. The AV is expected to overtake the static NPC vehicle to continue the journey. However, it remains stationary and fails to overtake (violating Article \#38).  Lastly, it may fail to make a necessary lane change and never reach the destination. In this situation, we set a destination that can be reached by a lane change after crossing the intersection ahead. However, the AV plans an unusual route, and keeps looping around and never reaches the destination. \\

\noindent \emph{\textbf {RQ3: How efficient is LawBreaker at generating test cases?}} Since there is no existing framework to support the evaluation of traffic laws, we compare our fuzzing algorithm against a fuzzing algorithm based on random generation. 
For our fuzzing algorithm, we set the initial population to 20 and the number of generations to 20, resulting in 420 test cases in one run. 
For random generation, we randomly generate 420 tests cases for each run. We run our fuzzing algorithm and random generation four times to reduce the effect of randomness, and the results are summarised in Table~\ref{tab:violation-coverage}. We evaluate \emph{Apollo6.0} and \emph{Apollo5.0} under all the testable traffic laws shown in Table~\ref{tab:coverage_of_failures}. According to the definition of $\Theta(\Phi)$ in Section~\ref{sec:sepcification_sub_violation_defination}, there are 82 possible violations. We compare our fuzzing algorithm with random generation in three different scenarios. Overall, we provide ten AVUnit scenario scripts at~\cite{ourweb}, and use three of the scenarios for fuzzing because they are common real-world scenarios that happen to be associated with complex traffic laws (e.g.~vehicle behaviour at junctions, overtaking).

\begin{itemize}
    \item \emph{S1}: In this scenario, there is a T-junction with traffic lights ahead. There are four NPC vehicles in the scenario. The ego vehicle is expected to cross the intersection safely.
    \item \emph{S2}: In this scenario, there are a few static and low-speed NPC vehicles ahead, and the ego vehicle is expected to overtake these static vehicles to reach the destination. There are five NPC vehicles in the scenario.
    \item \emph{S3}: In this scenario, there is an intersection with traffic lights ahead and the lanes are of two opposite directions. There are five NPC vehicles in the scenario. 
\end{itemize}

As can be seen from Table~\ref{tab:violation-coverage},
our fuzzing algorithm outperforms random generation with respect to both versions of Apollo, showing its utility for automatic testing. Furthermore, when comparing driving strategies, \emph{Apollo6.0} is more inclined to aggressive ones than \emph{Apollo5.0}, leading to more violations of traffic laws for  \emph{Apollo6.0}. For the four runs of the above three scenarios, we generate 77 scenarios that can cause accidents for \emph{Apollo5.0} and 96 for \emph{Apollo6.0}. 
As mentioned before, the implementation of deep learning for the decision process of \emph{Apollo6.0} is the main reason for this difference.\\

\begin{table}[]
\footnotesize
\centering
\caption{Violation coverage of different drivers}
\label{tab:violation-coverage}
\begin{tabular}{>{\centering}m{0.07\linewidth}ccccccc}
\hline
Scenario & Driver & Alg. & R1 & R2 & R3 & R4 & Avg \\\hline
&  &  Ours  &  27/82  &  25/82  &  21/82  & 27/82  & \textbf{25} \\ 
\multirow{-2}{*}{S1} & \multirow{-2}{*}{Apollo6.0}                                              
& Rand  &  26/82  &  23/82  &  15/82  & 21/82  & \textbf{21.25}    \\ \hline
    
&   & Ours   &  23/82  &  24/82  &  26/82  & 27/82  & \textbf{25} \\ 
\multirow{-2}{*}{S2} & \multirow{-2}{*}{Apollo6.0}                                              
& Rand  &  22/82  &  22/82  &  15/82  & 22/82  & \textbf{20.25} \\ \hline

&   & Ours  &  24/82  &  22/82  &  25/82  & 23/82  & \textbf{23.5} \\ 
\multirow{-2}{*}{S3} & \multirow{-2}{*}{Apollo6.0}                                              
& Rand  &  15/82  &  15/82  &  23/82  & 22/82  & \textbf{18.75}    \\ \hline


&  &  Ours  &  27/82  &  22/82  &  22/82  & 23/82  & \textbf{23.5} \\ 
\multirow{-2}{*}{S1} & \multirow{-2}{*}{Apollo5.0}                                              
& Rand  &  22/82  &  21/82  &  21/82  & 21/82  & \textbf{21.25}    \\ \hline
    
&   & Ours   &  17/82  &  16/82  &  17/82  & 15/82   & \textbf{16.25} \\ 
\multirow{-2}{*}{S2} & \multirow{-2}{*}{Apollo5.0}                                              
& Rand  &  15/82  &  15/82  &  14/82  & 15/82  & \textbf{14.75} \\ \hline

&   & Ours  &  25/82  &  24/82  &  24/82  & 25/82  & \textbf{24.5} \\ 
\multirow{-2}{*}{S3} & \multirow{-2}{*}{Apollo5.0}                                              
& Rand  &  25/82  &  23/82  &  24/82  & 23/82  & \textbf{23.75}    \\ \hline

\end{tabular}
\end{table}

\noindent \emph{\textbf {Threats to Validity. }}
Due to the nature of simulation-based testing, there are threats to the validity of the discovered issues.
For instance, some issues may only occur because of the latency between the simulator and the ADS. 

To solve this problem, Apollo itself has some built-in mechanisms to handle these situations such as the “estop” command to stop the vehicle.
Moreover, we have the following strategies to reduce the false-positive rate. First, all the found issues are repeated at least three times to ensure reproducibility. 
Second, the information we use for specifications is exactly the same as the ADS gets. In this way, even when there is a delay which causes the problem, we do not blame the ADS for it.
Third, we make sure the devices for simulations are in good condition~(e.g.~well-connected, sufficient memory).
Despite these measures, in general, we cannot rule out that a discovered problem may be due to the simulator.
Nonetheless, uncovering such a problem may still be helpful for improving the system as a whole.

%% file: RelatedWork.tex
\section{Related Work} \label{sec:related}

\noindent \textbf{\emph{Critical Scenario Generation.}}
A scenario for AV testing consists of static parameters (e.g.~time, weather) and dynamic parameters (e.g.~trajectory of vehicles and pedestrians). The main goal of existing works about AV testing is to generate scenarios that can expose vulnerabilities of AVs. We divide existing works into two groups: \emph{recreating real-world scenarios}, and \emph{generating new scenarios}. 

The first group of works explores how to recreate real-world scenarios.
TNO~\cite{paardekooper2019automatic} provides a dataset containing 6000 kilometers of driving on public roads and promotes the development of scenario-mining algorithms.
AC3R~\cite{gambi2019generating}, and DEEPCRASHTEST~\cite{Bashetty20DeepCrashTest} reconstruct car crashes to evaluate ADSs based on the scenario data from the police reports and accident videos respectively.
K-medoids~\cite{nitsche2017pre} focuses on recreating scenarios at T-and four-legged junctions based on the recordings of junction crashes in the UK.
Recreating unsafe cut-ins based on human driver lane change behaviour is another way of accelerating the evaluation of AVs~\cite{zhao2017accelerated}.
Extracting features of real-world scenarios to evaluate ADSs by comparing them with human drivers is also a solution~\cite{roesener2016scenario}.

The second group of works explores how to generate critical scenarios and defines the criticality of the scenario differently. 
AVFuzzer~\cite{li2020av} proposes an autonomous way to generate critical scenarios by fuzzing. The fuzzing algorithm of AVFuzzer is optimised with respect to (only) the distance from other NPC vehicles, i.e~looking for scenarios that are likely to cause collisions. 
NADE~\cite{feng2021intelligent} automatically generates scenarios that are natural and critical at the same time based on the data collected from a real-world dataset.
The criterion for evaluating whether a scenario is critical or not in NADE is the distance from other vehicles.
Similarly, `no collision' is also the criteria of Rule-based Searching~\cite{masuda2018rule}, Evolutionary-Algorithm-based Generation ~\cite{klischat2019generating}, and CMTS~\cite{ding2020cmts}.
A few works explore critical criteria beyond `no collision'.
Hungar~\cite{hungar2017test}, for example, defines the criticality of scenarios using a calculation over several harmful events-related variables.
The Baidu group proposes a coverage-based feedback mechanism~\cite{https://doi.org/10.48550/arxiv.2106.00873} that takes the coverage of the driving area of the map as the criterion, i.e.~covering a larger driving area indicates a better scenario.
Hauer and Schmidt~\cite{hauer2019did} explore how to automatically or manually generate test cases that cover all categories and model this problem as a Coupon Collector’s problem.
PlanFuzz~\cite{wan2022too} focuses on overly-conservative ADS behaviours and checks whether the ego vehicle stops in safe conditions.
Mullins defines a `near-miss' by whether or not the scenario causes the AV to be at the boundary between distinct performance modes~\cite{mullins2017automated, mullins2018accelerated}.
Althoff and Lutz~\cite{althoff2018automatic} define the criticality of the scenario by the size of the passable area, and generate critical scenarios by minimising the area.
Beglerovic et al.~\cite{beglerovic2017testing} induce a cost function from the specification of scenarios.

Although existing works propose different methods to generate scenarios for the evaluation of AVs, they focus on weak oracles and they lack a set of systematic time-tested oracles for AV testing. In this work, we evaluated AVs under traffic laws and propose the \coolname fuzzing algorithm to generate `critical' scenarios that are likely to violate the traffic laws in different ways. \\

\noindent \textbf{\emph{Formalisation of Specifications.}}
Existing works on robotic motion plan have implemented STL for describing complex oracles, e.g.~\cite{fainekos2009temporal,fainekos2005temporal, kress2009temporal,kress2007s,lahijanian2010motion,lahijanian2011temporal, fainekos2011revising,guo2013revising,kundu2019energy,shoukry2017linear,liu2017communication}. These works demonstrate the relevance of STL for motion-related specifications. 
Existing STL-based AV-related specification languages, e.g.~\cite{dreossi2019verifai,tuncali2019requirements,AVUnit2021, RSS-Tool}, cannot describe traffic laws.
Although AVUnit~\cite{AVUnit2021}, VERIFAI~\cite{dreossi2019verifai}, and Sim-ATAV~\cite{tuncali2019requirements, RSS-Tool} propose concrete ways to write specifications for evaluating ADSs,
they do so in a global perspective which is not suitable for specifying traffic laws in a driver-oriented manner.
Some existing works~\cite{rizaldi2015formalising, esterle2020formalizing, maierhofer2020formalization, rizaldi2017formalising, censi2019liability, collin2020safety} provide some formalisation methods for traffic laws. Rulebook~\cite{censi2019liability, collin2020safety} describes traffic laws by connecting atomic rules. The four formalisation methods~\cite{rizaldi2015formalising, esterle2020formalizing, maierhofer2020formalization, rizaldi2017formalising} propose different ways to describe traffic laws, but are limited to specific traffic laws
and do not provide a general driver-oriented style of language. 
For instance, none of these formalisation methods consider a number of important ego vehicle signals~(e.g.~high/low beam, left/right turn signal) and traffic signals~(e.g.~traffic lights, stop sign).
Moreover, their specifications must be customised for different test scenarios, and lack an automatic method to generate new scenarios as \coolname's optimising fuzzing algorithm does.

In this paper, we proposed the STL-based \coolname to translate traffic laws in a fully decoupled manner, i.e.~without any knowledge of the underlying scenario.

%% file: Conclusion.tex
\section{conclusion and future work} \label{sec:conclude}

In this paper, we proposed a framework, \coolname, for the evaluation of AVs with respect to road traffic laws. A key contribution is its driver-oriented specification language for the description of traffic laws, which is fully decoupled from and compatible with different scenario description DSLs.
We proposed a fuzzing engine that searches for different ways of violating the law specifications by maximising a form of specification coverage, i.e.~different ways of violating the underlying STL formulas.
We implemented and evaluated \coolname for the state-of-the-art Apollo ADS and LGSVL simulator, and were able to violate 14 Chinese traffic laws, with 173 of the generated test cases causing accidents.

There are several interesting avenues for future work. First,
we are interested in finding more efficient methods to generate test cases based on the given specification.
Furthermore, we only considered how the ego vehicle should behave in this work, and are interested in exploring how the traffic flow should be when other traffic participants are autonomous vehicles as well.